\documentclass[aps,prl,letterpaper,twocolumn,floatfix]{revtex4}
\usepackage{graphicx}

\newcommand{\del}{\mbox{\boldmath{$\nabla$}}}
\newcommand{\nhat}{\mbox{\boldmath{$\hat{n}$}}}

\newcommand{\xhat}{\mbox{\boldmath{$\hat{x}$}}}

\newcommand{\phib}{\mbox{\boldmath{$\phi$}}}
\newcommand{\re}{\mbox{Re\hspace{1pt}}}
\newcommand{\im}{\mbox{Im\hspace{1pt}}}

\newcommand{\Fb}{\mbox{{\bf F}}}
\newcommand{\Qb}{\mbox{{\bf Q}}}
\newcommand{\ub}{\mbox{{\bf u}}}
\newcommand{\vb}{\mbox{{\bf v}}}
\newcommand{\erf}{\mbox{erf\hspace{2pt}}}
\newcommand{\Pe}{\mbox{P\hspace{-1pt}e\hspace{1pt}}}
\newcommand{\Peo}{\mbox{P\hspace{-1pt}e}_o}

\begin{document}

\title{ Dynamics of Conformal Maps for a 
Class of Non-Laplacian Growth Phenomena }

\author{Martin Z. Bazant$^{1,2}$, Jaehyuk Choi$^1$, and Benny
Davidovitch$^3$}
\affiliation{$^1$ Department of Mathematics,
Massachusetts Institute of Technology, Cambridge, MA 02139 \\
$^2$  \'Ecole Sup\'erieure de
  Physique et de Chimie Industrielles, 10 rue Vauquelin, 75231 Paris,
  France\\
$^3$ ExxonMobil Research and Engineering,
 Route 22, Annandale, NJ 08801 }

%\author{Martin Z. Bazant}
%\email{bazant@mit.edu}
%\affiliation{ Department of Mathematics,
%Massachusetts Institute of Technology, Cambridge, MA 02139}
%\affiliation{
%\'Ecole Sup\'erieure de
%  Physique et de Chimie Industrielles, 10 rue Vauquelin, 75231 Paris,
%  France.}
%\author{Jaehyuk Choi}
%\affiliation{ Department of Mathematics,
%Massachusetts Institute of Technology, Cambridge, MA 02139}
%\author{Benny Davidovitch}
%\affiliation{ ExxonMobil Research and Engineering,
% Route 22, Annandale, NJ 08801}

% HISTORY:
% Oct 28, 2002:  MZB began this draft, based on BD early draft.
% Dec 2002 - Jan 2003: MZB revisions.
% Feb 7, 2003: MZB revisions based on meeting with JC and BD
% Feb 18, 2003: BD revisions based on meeting with MZB and JC
% Feb 25, 2003: MZB + JC revisions
% Mar 1, 2003: MZB revision, Austin, TX
% Mar 3, 2003: BD, JC comments, MZB revision 
% Mar 11: JC, BD revision
% Mar 12: MZB final draft.

\date{March 12, 2003}

\begin{abstract}
Time-dependent conformal maps are used to model a class of growth phenomena
limited by coupled non-Laplacian transport processes, such as nonlinear
diffusion, advection, and electro-migration. Both continuous and stochastic
dynamics are described by generalizing conformal-mapping techniques for
viscous fingering and diffusion-limited aggregation, respectively. A
general notion of time in stochastic growth is also introduced. The theory
is applied to simulations of advection-diffusion-limited aggregation in a
background potential flow. A universal crossover in morphology is observed
from diffusion-limited to advection-limited fractal patterns with an
associated crossover in the growth rate, controlled by a time-dependent
effective P\'eclet number. Remarkably, the fractal dimension is not
affected by advection, in spite of dramatic increases in anisotropy and
growth rate, due to the persistence of diffusion limitation at small scales.
\end{abstract}

\maketitle

Laplacian growth models describe some of the best known phenomena of
pattern formation far from equilibrium, including continuous dynamics
such as viscous fingering~\cite{bensimon86} and (quasi-static)
dendritic solidification ~\cite{kessler88} and stochastic processes
such as diffusion-limited aggregation (DLA)~\cite{witten81} and
dielectric breakdown~\cite{niemeyer84}.  In this class of models, the
interfacial velocity is determined by the normal derivative of a
harmonic function, so the powerful technique of conformal mapping has
been used extensively in two dimensions. Time-dependent conformal maps
are used in the classical analysis of continuous Laplacian
growth~\cite{polub45,shraiman84,feigenbaum01}, and the
analogous method of iterated conformal maps has recently been
developed for stochastic Laplacian growth~\cite{hastings98,david99,hastings01}.

In spite of the broad relevance of these models, real growth phenomena
often involve non-Laplacian transport processes, such as advection or
electro-migration coupled to diffusion~\cite{bouissou89,argoul95}. Much less
theoretical work exists in such cases, in part because conformal mapping
would appear to be of little use for non-harmonic functions.  An exception
is the recent use of streamline coordinates for dendritic solidification in
a potential flow, but it turns out that advection has no effect on the
shape of an infinite dendrite~\cite{cummings99}. Stochastic conformal-map
dynamics, however, has not yet been formulated for any non-Laplacian
transport process (although iterated conformal maps have been used in a
recent model of brittle fracture with a bi-harmonic elastic
potential~\cite{levermann02}).

In this Letter, we formulate the dynamics of conformal maps for growth
phenomena limited by non-Laplacian transport processes in a recently
identified conformally invariant class~\cite{bazant03}. We consider both
continuous and stochastic growth from a finite seed to allow nontrivial
competition between different transport processes. To illustrate the
theory, we study fractal growth driven by advection-diffusion in a
potential flow.

{\it Transport Processes --- } Consider a set of scalar `fields', $\phib =
\{ \phi_1, \phi_2, \ldots, \phi_M\}$, whose gradients produce quasi-static,
conserved `flux densities',
\begin{equation}
\Fb_i  = \sum_{j=1}^{M} C_{ij} \del \phi_j  \ , \ \ \
 \ \del\cdot\Fb_i = 0   \label{eq:Fi}
\end{equation}
where the coefficients, $\{C_{ij}(\phib)\}$, may be nonlinear functions of
the fields. This general system contains a number of physical
cases~\cite{bazant03}: ($M=1$) simple nonlinear diffusion, $\Fb_1 = - D(c)
\del c$, where $c$ is a temperature or particle concentration and $D(c)$
the diffusivity; ($M=2$) advection-diffusion, $\Fb_1 = - D(c) \del c + c
\del\phi $, of a scalar $c$ in a potential flow, $\ub = \Fb_2 =\del \phi$,
as well as ionic transport in a supporting electrolyte; and ($M\geq 2$)
ionic transport in a neutral electrolyte, $\Fb_i = -D_i(c_i) \del c_i -
b_i(c_i) q_i c_i \del \phi $, where $c_i$, $D_i$, $b_i$, and $q_i$ are
respectively the concentration, diffusivity, mobility, and charge of ion
$i$, and $\phi$ is the (non-harmonic) electrostatic potential, determined
implicitly by electro-neutrality, $\sum_{i=1}^M q_i c_i = 0$.

In planar geometries, it is convenient to represent a vector, $\Fb =
(F_x,F_y)$, as a complex scalar, $F = F_x + i F_y$, so
Equation~(\ref{eq:Fi}) takes the form,
\begin{equation}
F_i  = \sum_{j=1}^{M} C_{ij}
\nabla \phi_j  \ , \ \
 \ \re(\overline{\nabla} \Fb_i) = 0   \label{eq:Fic}
\end{equation}
in the $z=x+iy$ plane, where $\nabla = \frac{\partial}{\partial x} + i
\frac{\partial}{\partial y}$.
% (since ${\bf a \cdot b} = \re (\overline{a}b)$).
Under a conformal mapping, $w = f(z)$, $F_i$ transforms as 
\begin{equation}
F_i(z,\bar{z}) = \overline{f^\prime(z)}\, F_i(w,\overline{w}) .
\label{eq:complexFitrans}
\end{equation}
(like $\nabla$ ~\cite{needham}), and $\re ( \overline{\nabla} F_i ) =
0$ is unchanged. Therefore, even though the solutions (depending on
$z$ and $\bar{z} = x-iy$) are not harmonic functions, the usual trick
of conformal mapping still works~\cite{bazant03}: If
$\phib(w,\bar{w})$ solves Eq.~(\ref{eq:Fic}) in one domain,
$\Omega_w$, then $\phib(f(z),\overline{f(z)})$ solves
Eq.~(\ref{eq:Fic}) in another domain, $\Omega_z = f^{-1}(\Omega_w)$
(with appropriately transformed boundary conditions).

Interfacial dynamics in the plane can be elegantly described by a
conformal map, $z=g(w,t)$, from $\Omega_w$, the exterior of the unit
circle, to $\Omega_z(t)$, to the exterior of the (singly connected)
growing
object~\cite{polub45,shraiman84,feigenbaum01,hastings98,david99,hastings01}. Since
the map must be univalent (1-1), it has a Laurent series,
\begin{equation}
g(w,t) = A_1(t) w + A_0(t) + \frac{A_{-1}(t)}{w} + \ldots, \ \ \ |w|>1 \ ,
\label{eq:laurent}
\end{equation}
where $A_1(t)$ is real and defines an effective diameter of
$\Omega_z(t)$~\cite{hastings98}. As described above, the fields
satisfying~(\ref{eq:Fi}) in $\Omega_z(t)$ are easily obtained from
the inverse, $w = f(z,t)$, once the same equations are solved in
$\Omega_w$. As in Laplacian growth, the removal of geometrical
complexity from the transport problem is a tremendous simplification.

{\it Boundary Conditions ---} (BC) Along the moving boundary, $\partial
\Omega_z(t)$ we consider generalizations of Dirichlet ($\phi_i=0$) and
Neumann ($\nhat\cdot\del \phi_i = 0 $) BC ~\cite{bazant03}:
\begin{equation}
R_i\left(\phib(z,\overline{z})\right)=0 \ \  \mbox{or} \ \ 
\nhat\cdot\Fb_i = \re \left(\overline{n(z)} F_i(z,\overline{z})\right) = 0
, \label{eq:bc}
\end{equation}
respectively, for $z \in \partial\Omega_z(t)$, where $n = n_x+ i n_y$
represents the outward normal, $\nhat$, and $R_i(\phib)$ is a function of
the fields. The former BC express interfacial equilibrium for `fast
reactions' (compared to transport rates), while the latter expresses
impermeability to flows ($\Fb = \ub = \del\phi$) or flux densities of
non-reacting species. Due to Eq.~(\ref{eq:complexFitrans}), these BC
are the same,
\begin{equation}
R_i\left(\phib(w,\overline{w})\right)=0 \ \ \ \mbox{or}  \ \ \
\nhat\cdot\Fb_i = \re \left(\overline{w} F_i(w,\overline{w})\right) = 0  ,
\label{eq:bcw}
\end{equation}
for $w \in \partial\Omega_w$, since $n(w) = w = n(z)
f^\prime(z)/|f^\prime(z)|$.

Far from the growth, we assume either constant values of the fields (e.g.
temperature, concentration) or given flux (or flow) profiles which drive
the growth:
\begin{equation}
\phi_i(z,\bar{z}) \to \phi_i^{\infty}   \ \ \  {\mbox{or}}\  \ \  
F_i(z,\bar{z}) \sim F_i^\infty(z,\bar{z}) \label{eq:bcinf}
\end{equation}
as $|z|\to\infty$ (where $F_i^\infty$ could also vary in time). The former
BC also remain the same after conformal mapping, but the latter is
transformed by Eqs.~(\ref{eq:complexFitrans}) and (\ref{eq:laurent}):
\begin{equation}
\phi_i(w,\bar{w}) \to \phi_i^{\infty} \  \ \ \ {\mbox{or}} \ \ \  \
F_i(w,\bar{w}) \sim \overline{A_1} F_i^\infty(A_1 w,\overline{A_1 w})
\label{eq:bcinfw}
\end{equation}
as $|w|\rightarrow\infty$. Through $A_1(t)$ $(=\overline{A_1}(t))$, the
fields and fluxes in $\Omega_w$ vary with the diameter of the growth,
$\Omega_z(t)$.

{\it Continuous Dynamics ---} Suppose that a Lagrangian boundary point,
$z(t) \in \partial \Omega_z(t)$, moves in the normal direction with
(complex) velocity,
\begin{equation}
v = z_t = \alpha\, n \, \sigma, \ \ \ \sigma = \re (\overline{n} Q)), \ \ \
Q = \sum_{i=1}^M B_i F_i ,  \label{eq:sigma}
\end{equation}
where $\Qb$ is a flux density causing growth, $\alpha$ is a constant, 
and $B_i(\phib)$ may be functions of the fields. This
generalizes Stefan's law ($\vb = \alpha\, \nhat\cdot\del\phi$), e.g. to
electrodeposition~\cite{argoul95} (where $\Qb$ is for the depositing ion).

The classical analysis of viscous fingering is easily generalized to
Eqs.~(\ref{eq:Fic}), (\ref{eq:bc}), (\ref{eq:bcinf}), and (\ref{eq:sigma}).
Let $w(t)$ be a `marker' for $z(t) = g(w(t),t)$ on $\Omega_w$
~\cite{feigenbaum01}. Substituting $z_t = g^\prime\, w_t + g_t$ into
Eq.~(\ref{eq:sigma}), multiplying by $\overline{w g^\prime}$, taking the
real part, and using $\re(\overline{w} w_t) = 0$ for $|w|^2 = \overline{w}w
= 1$, we arrive at an evolution equation for the conformal map,
\begin{equation}
\re ( \overline{ w \, g^\prime} \, g_t ) = \alpha \ \sigma(w,A_1(t)) \ \
\mbox{ for } \ \ |w|=1 \label{eq:gsb}
\end{equation}
where $\sigma(w,A_1(t)) = \re(\overline{w}Q(w,\overline{w}))$ is the normal
flux density on $\Omega_w$, which depends on $A_1(t)$ through
Eq.~(\ref{eq:bcinfw}).

The evolution equation for radial Laplacian growth (e.g. viscous
fingering)~\cite{polub45,shraiman84}, $\re(\overline{w \, g^\prime}\, g_t)
= 1$, corresponds to the special case of uniform flux, $\sigma=$ constant.
This dynamics is known to preserve number of pole-like singularities
(inside the unit circle) for a wide class of initial maps, $g(w,0)$. Except
for some elementary maps, e.g. circles and ellipses,
% ($g(w,0)=aw , \ aw + b/w$, respectively, with $a > 0\ , \  |b| < a/2$),
which preserve their shapes, arbitrary smooth initial interfaces
develop singularities (cusps) in finite time~\cite{shraiman84}.

In our generalized models, the time-dependent, non-uniform flux,
$\sigma$, in Eq.~(\ref{eq:gsb}) changes the analytic structure of an
initial map (e.g. number of poles), so even circles and ellipses
become distorted. This raises interesting questions about finite-time
singularities, e.g. What is the fate of solidification from a circular
seed in a flowing melt (with $\sigma$ described below)? Does advection
generally enhance or retard the formation of singularities? We leave
these questions for future work and focus here on non-Laplacian fractal
growth.

\begin{figure*}
\mbox{
\hspace{-0.1in}
\includegraphics[height=1.55in]{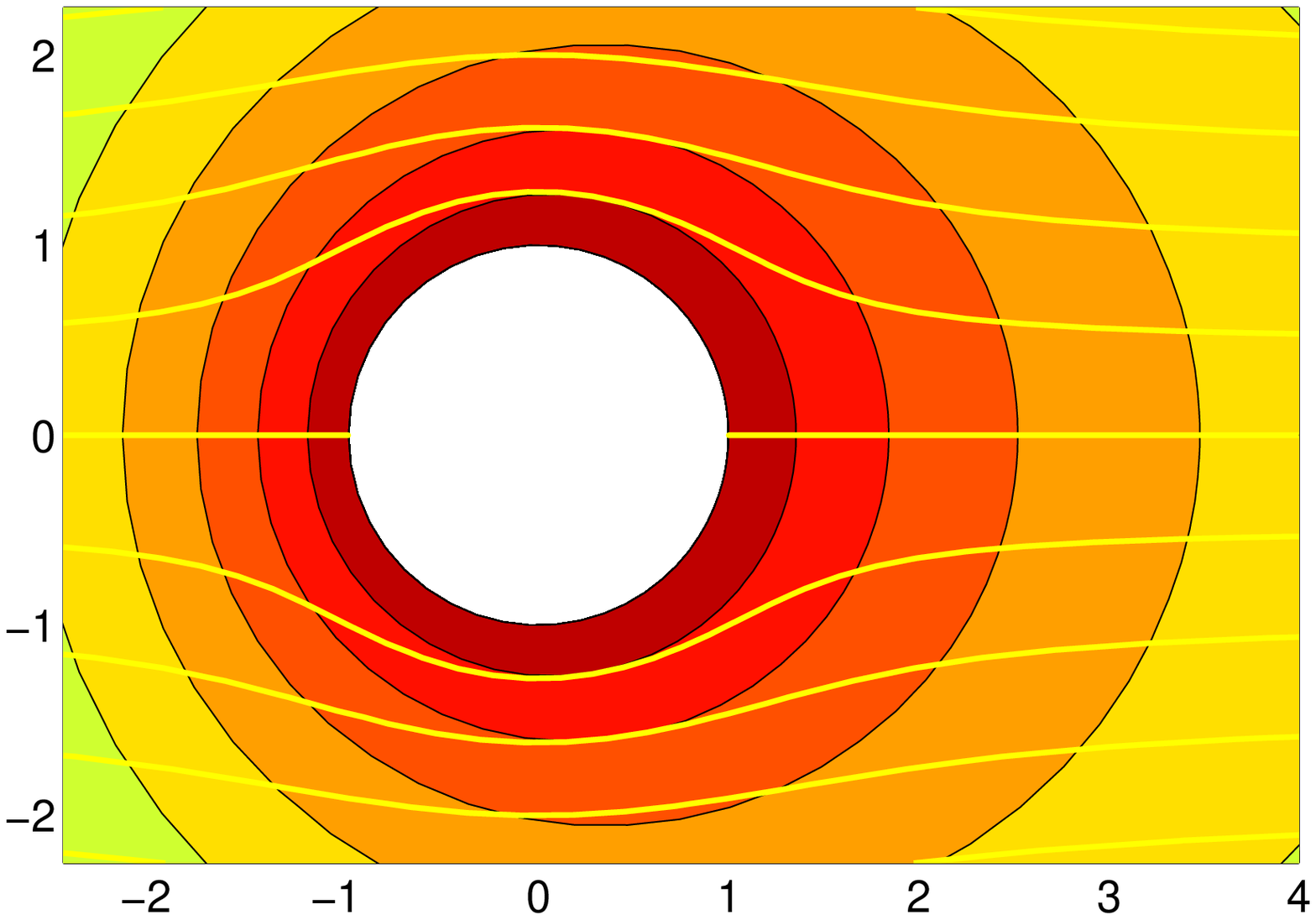}
} 
\nolinebreak \mbox{
\includegraphics[height=1.55in]{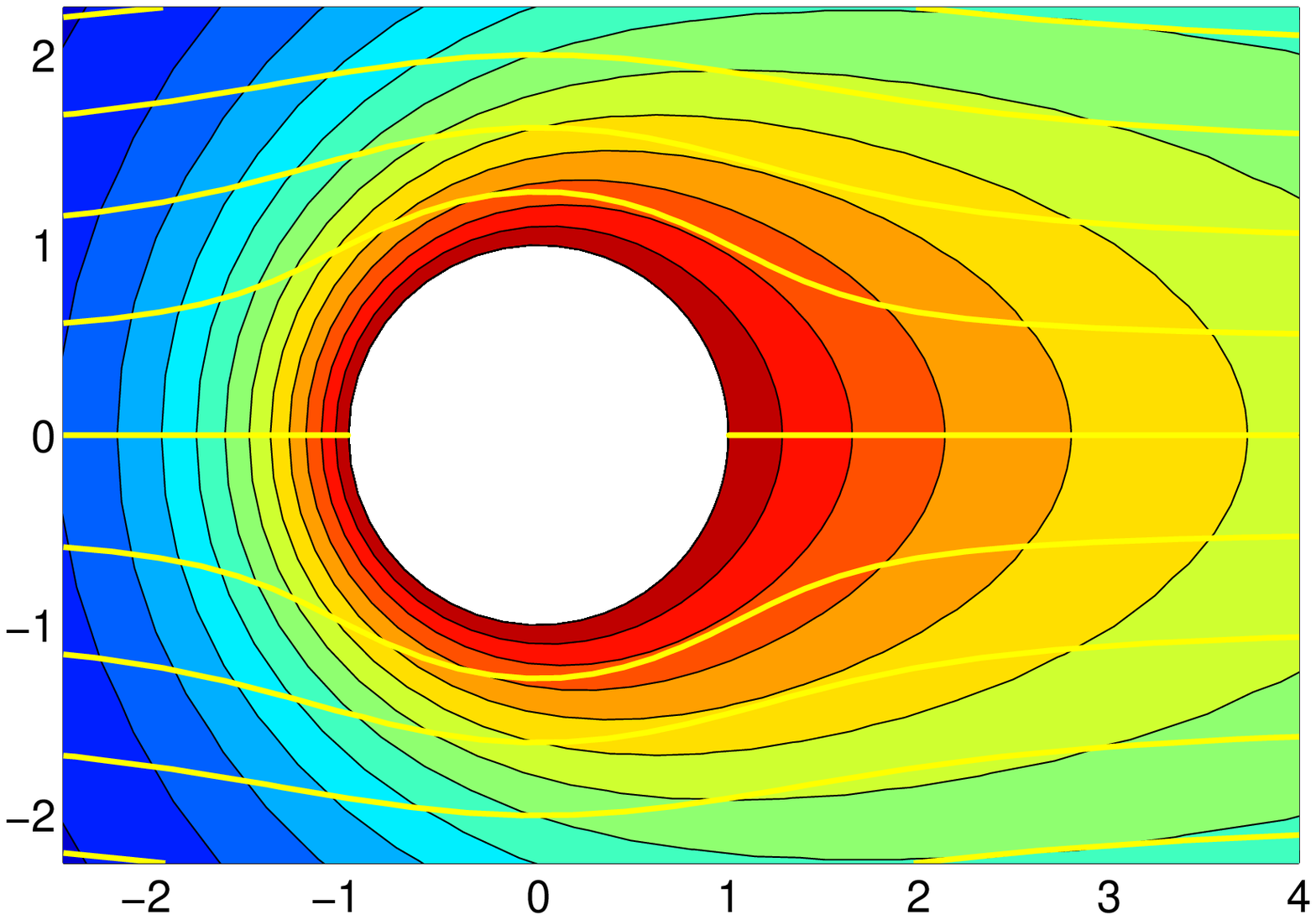}
} 
\nolinebreak \mbox{
\includegraphics[height=1.55in]{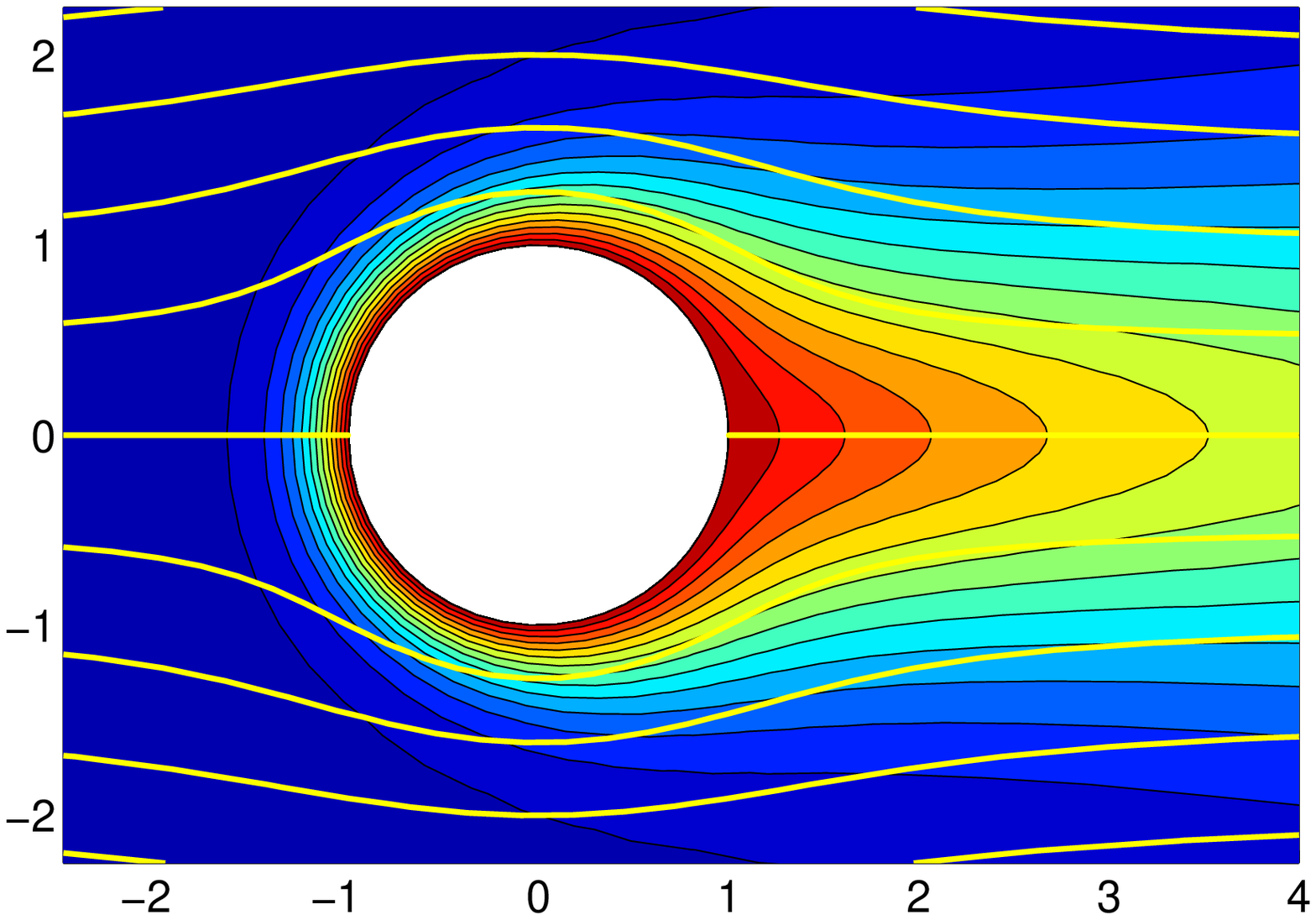}
} \\ 
\mbox{
\includegraphics[height=1.55in]{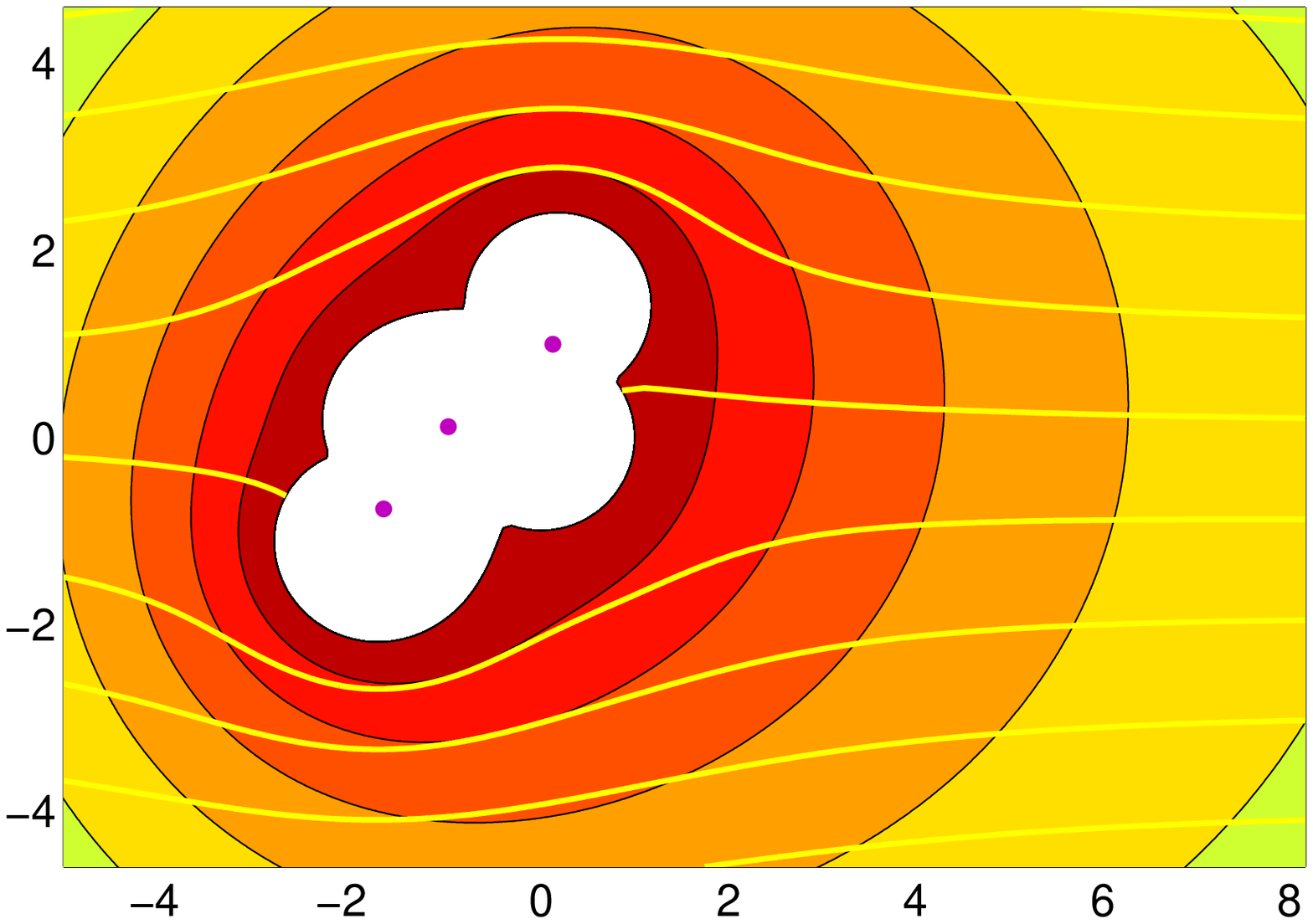}
} 
\nolinebreak 
\mbox{
\includegraphics[height=1.55in]{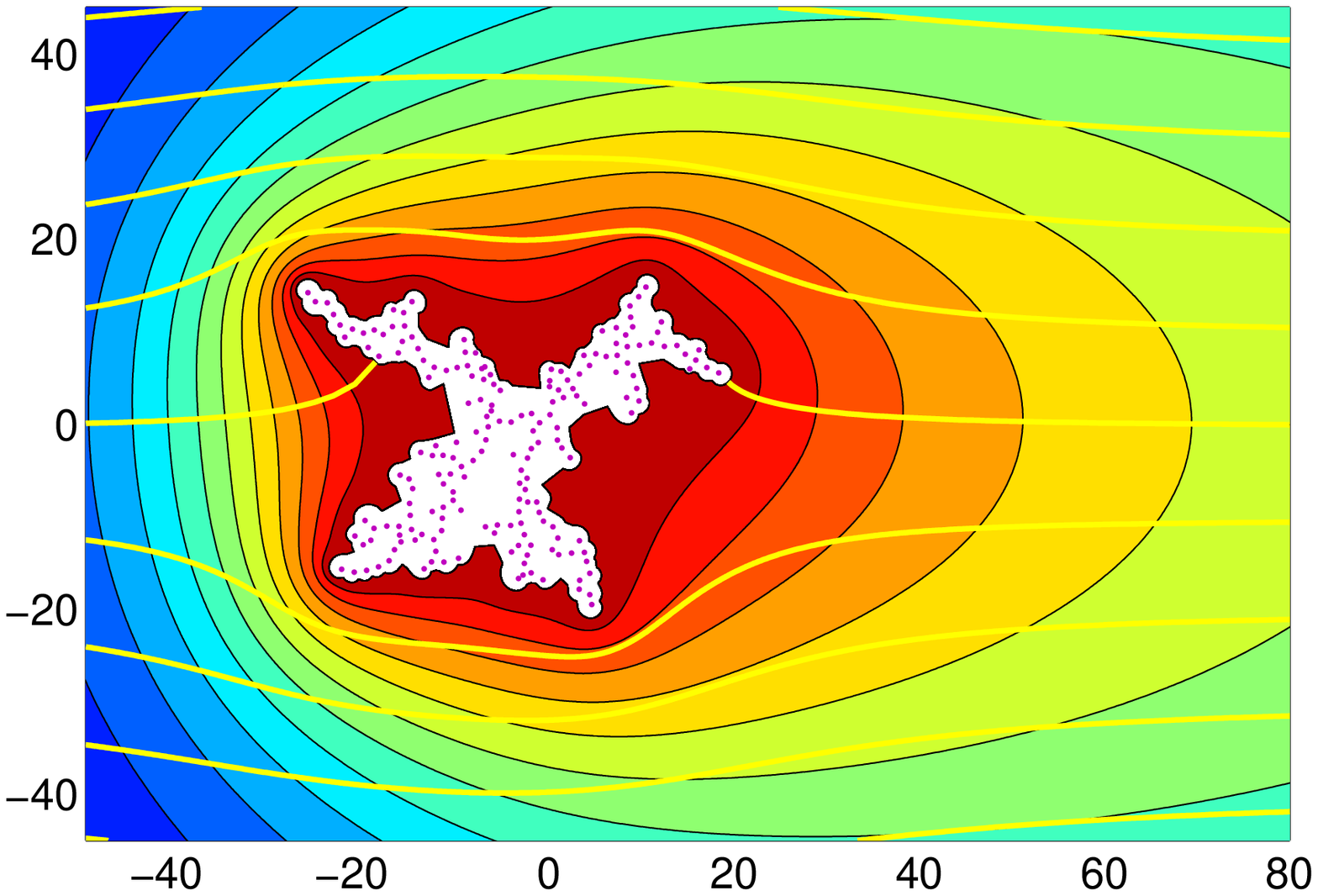}
} 
\nolinebreak 
\mbox{
\hspace{-0.13in}
\includegraphics[height=1.55in]{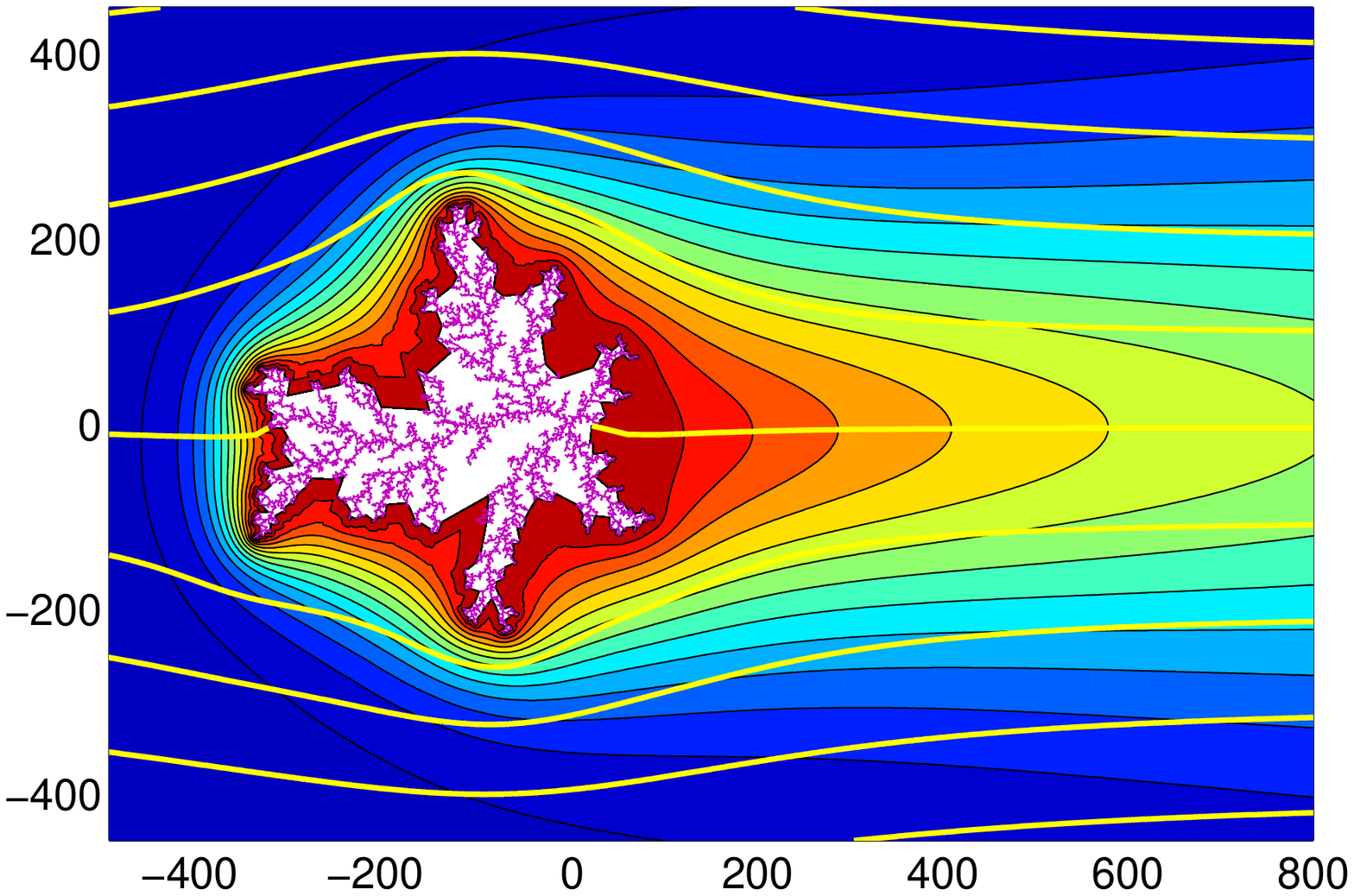}
} 
\caption{ Evolution of the flow (yellow streamlines) and concentration
(color contour plot) in $\Omega_w$ (top) and in $\Omega_z(t)$ (bottom)
during advection-diffusion-limited aggregation of $N=10,000$ particles
($\lambda_o= L_o^2=1$) with $\Peo=0.05$. The columns from left to
right correspond to $\Pe = 0.1$, $\Pe = 1$, and $\Pe = 10$. The lower
plots are shown rescaled by $A_1(t_3)=2$, $A_1(t_{193})=20$, and 
$A_1(t_{9621})=200$, the effective diameters of the fractal cluster.
\label{fig:fields} 
}
\end{figure*}

{\it Stochastic Dynamics ---} Suppose that the domain, $ \Omega_z(t_N)$,
grows from its initial shape, $\Omega_z(0)$, at times $t_1, t_2,\ldots,
t_N$ by discrete `bumps' representing $N$ particles of characteristic area,
$\lambda_o$ ~\cite{hastings98,david99}.  Since our models exhibit
non-trivial time-dependence (see below), we introduce time into the
usual morphological model by replacing 
Eq.~(\ref{eq:sigma}) with $p(z,t_N)= \alpha\, \sigma(z)\, / \lambda_o$,
for $z \in \partial\Omega_z(t_{N-1})$, where $p(z,t_N)\,|dz|\,dt$ is the
probability that the $N$th growth event occurs in the boundary element
$(z,z+dz)$ in the time interval $(t,t+dt)$. The waiting time,
$t_N-t_{N-1}$, is then an exponential random variable with inverse mean,
\begin{equation}
\tau_N^{-1} = \frac{\alpha}{\lambda_o} \oint_{\partial\Omega_z(t_{N-1})}
\hspace{-0.4in} \sigma(z)\,|dz| = \frac{\alpha}{\lambda_o} \int_0^{2\pi}
\sigma(e^{i\theta},A_1(t_{N-1})) \, d\theta ,  \label{eq:tauN}
\end{equation}
where we use $|dz| = |dw|/|f^\prime|$ and $\sigma(z) =
|f^\prime(z)|\sigma(w)$ from Eq.~(\ref{eq:complexFitrans}) to transform to
$\partial \Omega_w$ where $w=e^{i\theta}$ and $|dw|=d\theta$. The
probability that the position of the $N$th growth occurs in $(z,z+dz)
\subset \partial\Omega_z(t_{N-1})$ is, $p_N(z)|dz| =
\frac{\alpha}{\lambda_o} \,\tau_N\, \sigma(z)|dz| = P_N(\theta) d\theta$,
where
\begin{equation}
P_N(\theta) = \frac{\alpha}{\lambda_o} \, \tau_N \,
\sigma(e^{i\theta},A_1(t_{N-1})) , \ \ \ e^{i\theta} \in \Omega_w
\label{eq:PN}
\end{equation}
is the probability measure for angles on the unit circle. In
DLA~\cite{hastings98}, $p_N(z)$ is the harmonic measure;
$P_N(\theta)=(2\pi)^{-1}$ is uniform; and $\tau_N$ is not defined.

It is now straightforward to generalize the Hastings-Levitov DLA
algorithm~\cite{hastings98} to our non-Laplacian models. The univalent
map $z=g(w,t_N)$ from $\Omega_w$ to $\Omega_z(t_N)$, the exterior of
an $N$-particle aggregate, is constructed iteratively from a
two-parameter basic map that places a `bump' of area $\lambda_N$
around an angle $\theta_N$ on the unit circle. In order to fix the
area $\lambda_o$ of a new bump on $\partial \Omega_{z}(t_N)$, the size
of its pre-image, $\lambda_N$, on $\partial \Omega_{w}$ is divided by
the Jacobian of the previous map, $\lambda_k =
\lambda_o/|g^\prime(e^{i\theta_N},t_{N-1})|^2$. The first difference
with DLA is that the angle, $\theta_N$, is chosen according to the
time-dependent (non-harmonic) measure, $P_N(\theta)$, in
Eq.~(\ref{eq:PN}).  The second difference is the evolving waiting
time, $\tau_N$, in Eq.~(\ref{eq:tauN}).

{\it Advection-Diffusion-Limited Aggregation --- } (ADLA) As an example, we
consider the stochastic aggregation of particles around a circular seed of
radius, $L_o$, limited by advection-diffusion in a uniform potential flow
of speed $U$ and concentration $C$. In $\Omega_z(t)$ the transport problem
has the usual dimensionless form,
\begin{eqnarray}
\Peo \del \phi \cdot \del c = \del^2 c , \ \ \del^2 \phi = 0 ,  \ & &
  z \in \Omega_z(t) \label{eq:ad} \\ c=0 , \ \ \nhat\cdot\del\phi=0 ,
 \ \ \sigma = \nhat\cdot\del c, \  & &  z \in \partial\Omega_z(t)
 \label{eq:circbc} \\ c \to 1 , \ \ \del\phi
 \to \xhat, \  & &  |z| \to \infty \label{eq:infbc}
\end{eqnarray}
where $x$, $\phi$, $c$, and $\sigma$ are in units of $L_o$, $UL_o$, $C$,
and $D C/L_o$, respectively, and $\Peo = UL_o/D$ is the initial P\'eclet
number. In complex notation, we solve Eq.~(\ref{eq:Fic}) in $\Omega_z(t)$
with BC (\ref{eq:bc}) and (\ref{eq:bcinf}) for the `fluxes', $F_1 = -\Peo
\nabla c + c \nabla \phi$ and $F_2 = u = \nabla \phi$. In $\Omega_w$,
Equations~(\ref{eq:Fic}) and (\ref{eq:bcw}) are the same, but the BC
(\ref{eq:bcinf})  calls for a background flow speed at $|w|\to\infty$ that
diverges with $A_1(t)$.

It is natural then to rescale the $w$-velocity by $A_1(t)$ to fix the
background speed at unity, and instead solve the same equations in
$\Omega_w$ as in $\Omega_z(t)$ with a {\it time-dependent P\'eclet number},
$\Pe(t) = \Peo\, A_1(t)$. Since $A_1(t)$ is an effective diameter for
$\Omega_z(t)$, the theory has shown us how to properly define $\Pe$ (which
is not obvious for fractals). We also see that advection eventually
dominates diffusion since $\Pe(t) \rightarrow\infty$, so we expect to see a
crossover from DLA to new advection-limited dynamics, as shown in the
simulation of Fig.~\ref{fig:fields}, which we now discuss in detail.

With the scalings above, the velocity potential in $\Omega_w$ has the usual
harmonic form, $\phi(w) = \re(w + 1/w)$, for potential flow past a
cylinder, but the non-harmonic concentration, $c(w,\overline{w},\Pe(t))$,
cannot be expressed in terms of elementary functions. However, asymptotic
expansions (for fixed $w$) can be
derived~\cite{asympnote},
\begin{equation}
c(w,\overline{w},\Pe) \sim \left\{ \begin{array}{ll} -\re \left(\log w /
\log \Pe\right) &  \ \Pe \ll 1 \\ \erf\left[\sqrt{\Pe}\, \im\left(\sqrt{w}
+\frac{1}{\sqrt{w}}\right)\right] &  \ \Pe \gg 1
\end{array} \right.
\label{eq:clim}
\end{equation}
The low-$\Pe$ approximation, the familiar harmonic field of Laplacian
growth, is valid out to a `boundary layer at $\infty$', while the
high-$\Pe$ approximation is valid away from a thin wake on the
positive real axis (the branch cut for
$\sqrt{w}$)~\cite{bazant03}. From Eqs.~(\ref{eq:tauN}), (\ref{eq:PN}),
and (\ref{eq:clim}) we also obtain,
\begin{equation}
\tau_N \sim \left\{ \begin{array}{l} -\log \Pe  \\ 8 \sqrt{\pi/\Pe}
\end{array} \right. \ \ \ P_N(\theta) \sim \left\{ \begin{array}{ll}
1/2\pi & \ \Pe \ll 1 \\ \frac{1}{4} \sin \frac{\theta}{2} & \ \Pe \gg 1
\end{array} \right.   \label{eq:fixedpoints}
\end{equation}
where $\tau_N$ is measured in units of $(\lambda_o/D)/(\alpha C)$. Even for
$\Peo \ll 1$, the ADLA dynamics smoothly crosses over from the DLA
`unstable fixed point', $\Pe(t) \ll 1$,  to a new advection-dominated
`stable fixed point', $\Pe(t) \gg 1$, described by
Eq.~(\ref{eq:fixedpoints}).

%\begin{figure}
%\mbox{ \includegraphics[width=2.3in]{NormalFlux.eps} } \caption{ The
%probability measure, $P_N(\theta)$, (or normal flux
%$\sigma(e^{i\theta},\Pe)$)) on the unit circle, $\partial\Omega_w$, scaled
%to its maximum value (at the leading edge at $\theta=\pi$) for steady
%advection-diffusion to an absorbing cylinder in a uniform potential flow at
%$\Pe = 0, 2^{-8}, 2^{-6},...,2^2,\infty$. \label{fig:sigma} }
%end{figure}

To investigate this crossover, we accurately obtain the normal flux,
$\sigma(w,\Pe(t))$, on $\partial \Omega_w$ by interpolating {\it static}
numerical solutions for a range $\Pe$. Following the methods above, we then
perform many ADLA simulations of $N>10^4$ particles for various $\Peo$ and
$\lambda_o/L_o^2$. Details will be given elsewhere, but here we briefly
discuss morphological changes and growth rate.

\begin{figure}
\mbox{ \hspace{-0.1in} \includegraphics[width=1.65in]{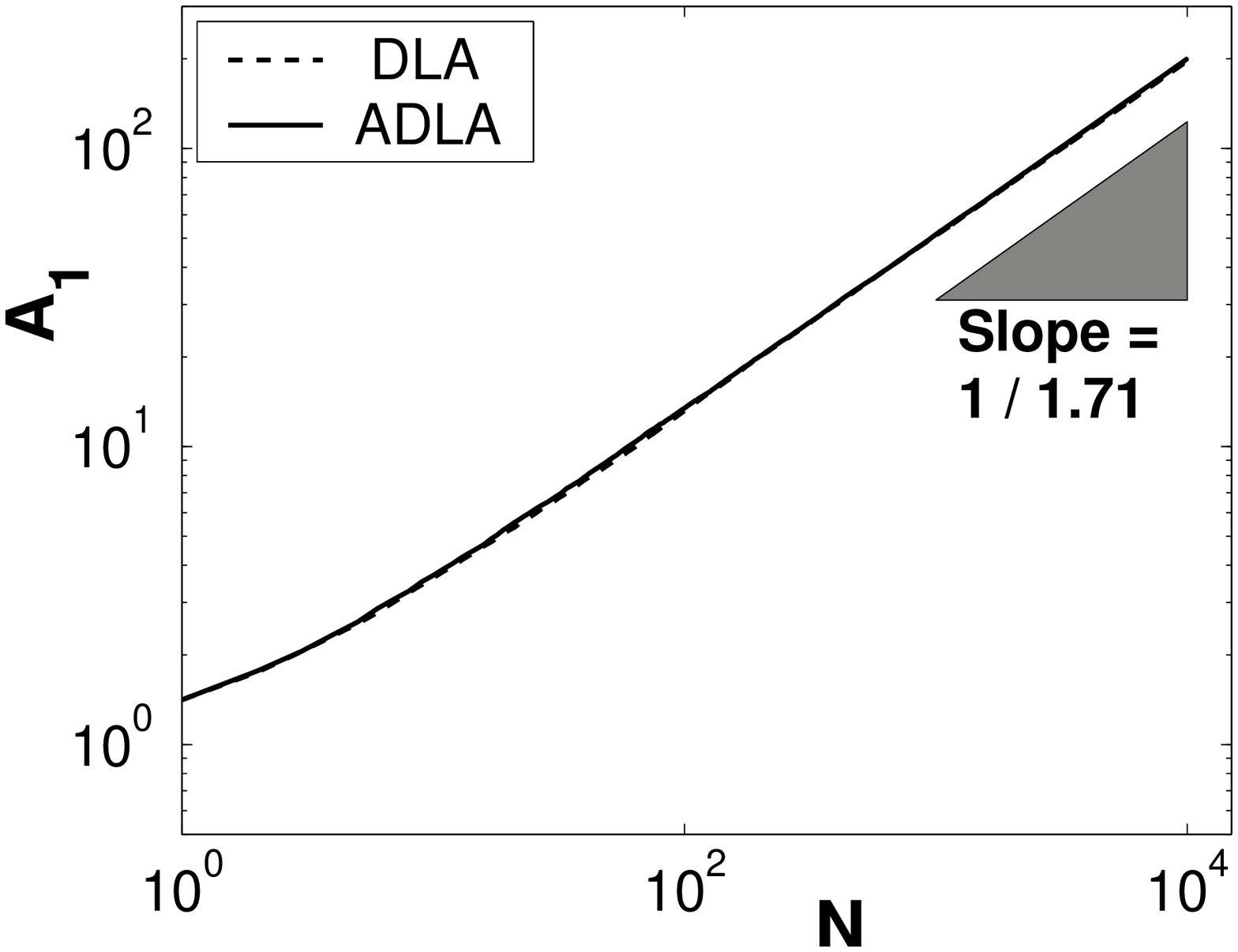} }\nolinebreak \mbox{
\includegraphics[width=1.65in]{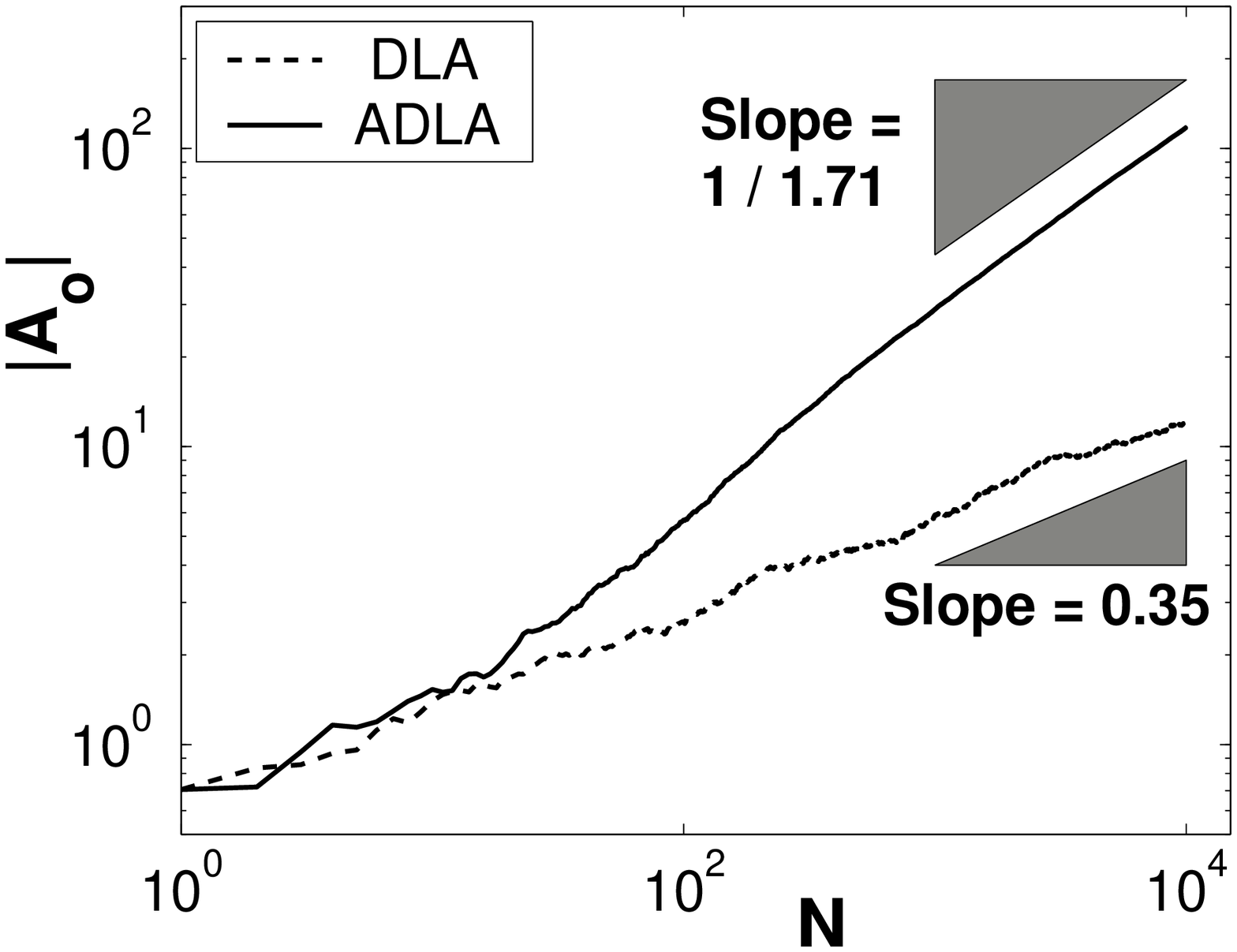} } \caption{ Log-log plot of
the Laurent coefficients, $A_1(t)$ and $|A_0(t)|$ versus $N(t)$, averaged
over 30 ADLA simulations ($\Peo = 0.05$, $\lambda_o = L_o^2 = 1$), compared
with analogous results for DLA without advection. 
 \label{fig:scalingA1A0} }
\end{figure}

The Laurent coefficients in Eq.~(\ref{eq:laurent}) contain morphological
information~\cite{hastings98,david99}. As illustrated in
Fig.~\ref{fig:scalingA1A0}, for all $\Peo$, the diameter of the cluster,
$A_1(t_N) \sim N^{1/D_f}$, remarkably maintains the same fractal dimension
as DLA, $D_f = 1.71$, for all $\Pe(t)$ and $N(t)$ in the scaling regime,
$A_1(t_N) \gg 1$. (As a check, we obtain the same $D_f$
from the radius of gyration.) {\it A posteriori}, this can be understood by
noting that the stable fixed point has a growth measure, $P_N(\theta)
\propto \sin \theta/2$, which is differentiable, and thus locally constant
(as in DLA), everywhere except at the rear stagnation point, $\theta=0$.
Physically, this simply means that diffusion always dominates advection at
small scales. We conjecture that the same $D_f$ holds for any non-Laplacian
dynamics in our class, if $\lim_{N\rightarrow\infty}P_N(\theta)$ is
continuous and almost everywhere differentiable. The surprising
universality of $D_f=1.71$ makes its exact value seem quite
fundamental.

Of course, as seen in Fig.~\ref{fig:fields}, the growth is highly
anisotropic and moves toward the flow for $\Pe(t) \gg 1$. This is seen in
scaling of the next Laurent coefficient, $A_0(t)$, the `center of
charge'~\cite{david99},
\begin{equation}
|A_0(t_N)| \sim \left\{ \begin{array}{ll} N^a & \ \ \ \Pe \ll 1 \\
N^{1/D_f} & \ \ \ \Pe \gg 1
\end{array} \right.
\end{equation}
which crosses over from DLA scaling ($2a = 0.7$ ~\cite{david99})
to the same scaling as $A_1(t_N)$, as shown in
Fig.~\ref{fig:scalingA1A0}. Elsewhere, we will show that
$|A_0(t)|/A_1(t)$ tends to a universal function of $\Pe(t)$ after
initial transients vanish, $A_1(t) \gg 1$, e.g.
$\lim_{t\to\infty}|A_0(t)|/A_1(t) = 0.6$.

%Recovering time from $\{\tau_N\}$ shows that 
The expected total mass versus time, $N(t)$, also undergoes a
crossover. Using Eqs.~(\ref{eq:tauN}) and (\ref{eq:fixedpoints}) and
integrating $dN/dt = \tau_N^{-1}(\Pe(t)) = \tau_N^{-1}(\Peo N^{1/D_f})$
yields, 
%
%$N(t) \sim t/|\log\Pe|$ for $\Pe\ll 1$ and $N(t)\sim
%(\frac{1}{8}\sqrt{\Pe/\pi}\;t)^{2D_f/(2D_f-1)}$ for $\Pe \gg 1$, 
%
$N(t) \propto t$ for $t \ll 1$ and $N(t)\propto t^{2D_f/(2D_f-1)}$
for $t \gg 1$, where again only $D_f$
matters. The cluster diameter $A_1(t)$ switches from $t^{1/D_f}$ to
$t^{2/(2D_f-1)}$ scaling in time, even though the scaling with
$N(t)$ does not change. One should also bear in mind that $\tau_N \to
0$ as $\Pe(t) \to \infty$, so the
quasi-static, discrete-growth approximation must eventually break down
(although this is delayed in the dilute limit, $\alpha C
\ll 1$).

In summary, we have formulated conformal-map dynamics for a class of
planar growth phenomena limited by non-Laplacian transport
processes. Although various simplifying assumptions were made, the
method enables very efficient simulations of fractal-growth phenomena,
such as ADLA, which it seems could not be achieved by more efficient way. By
describing competing transport processes, it also paves the way for
new studies of general crossover phenomena in pattern formation.

\vspace{0.1in} The authors would like to thank M. Ben-Amar, D. Margetis,
and T. M. Squires for helpful discussions and ESPCI for support through the
Paris Sciences Chair (MZB).

\nopagebreak


\begin{thebibliography}{99}
\bibitem{bensimon86} D. Bensimon, L. Kadanoff, B. I. Shraiman,
and C. Tang,
% Viscous flows in two dimensions
Rev. Mod. Phys. {\bf 58}, 977 (1986).

\bibitem{kessler88} D. A. Kessler, J. Koplik and H. Levine,
% Pattern selection in fingered growth phenomena
Adv. Phys. {\bf 37}, 255  (1988).

\bibitem{witten81} T. A.  Witten and L. M.  Sander
% DLA: A Kinetic critical phenomenon
Phys. Rev. Lett.  {\bf 47}, 1400 (1981).

\bibitem{niemeyer84} L. Niemeyer, L. Pietronero, and H. J. Wiesmann,
Phys. Rev. Lett. {\bf 52}, 1033 (1984).

\bibitem{polub45} P. Ya. Polubarinova-Kochina,
Dokl. Akad. Nauk. S. S. S. R. {\bf 47}, 254 (1945); \ L. A. Galin, {\bf
47}, 246 (1945).

\bibitem{shraiman84}  B. I. Shraiman and D. Bensimon,
Phys. Rev. A, {\bf  30}, 2840 (1984).

%\bibitem{howison92} S. D. Howison, Eur. J. Appl. Math. {\bf 3}, 209 (1992).

\bibitem{feigenbaum01} M. J. Feigenbaum, I. Procaccia, and B. Davidovitch,
J. Stat. Phys. {\bf 103}, 973 (2001).

\bibitem{hastings98} M. Hastings and L.  Levitov (1998) Physica D {\bf
116}, 244 (1998).

\bibitem{david99} B. Davidovitch, H. G. E. Hentschel, Z. Olami, I.
Procaccia, L. M. Sander, and E. Somfai, 
% DLA and iterated conformal maps
Phys. Rev. E {\bf 59}, 1368 (1999).

\bibitem{hastings01} M. B. Hastings, Phys. Rev. Lett. {\bf 87}, 175502 (2001).

\bibitem{bouissou89} P. Boussiou, B. Perrin, and P. Tabeling,
% Influence of an external flow on dendritic crystal growth
Phys. rev. A {\bf 40}, 509 (1989); \ Y.-W. Lee, R. Ananth, and W. N. Gill,
% Selection of a length scale in unconstrained denditic growth with
% convection in the melt
J. Crystal Growth {\bf 132}, 226 (1993).

\bibitem{argoul95}  
F. Argoul and A. Kuhn, Physica A {\bf 213}, 209 (1995); 
% The influence of transport and reaction processes on the morphology
% of a metal electrodeposit in thin gap geometry (pp. 209-231).
\ J. M. Huth, H. L. Swinney, W. D. McCormick, A. Kuhn, and F. Argoul, Phys. Rev. E
{\bf 51}, 3444 (1995).
%         ROLE OF CONVECTION IN THIN-LAYER ELECTRODEPOSITION

%\bibitem{boussinesq1905} M. J. Boussinesq, J. Math. Pure. Appl. {\bf
%1} 285 (1905).

\bibitem{cummings99} L. M. Cummings, Y. E. Hohlov, S. D. Howison, and
K. Kornev,
% Two-dimensional solidification and melting in potential flows
J. Fluid Mech. {\bf 378}, 1 (1999).

\bibitem{levermann02} F. Barra, H. G. E. Hentschel, A. Levermann,
and I.  Procaccia,
% Quasistatic fractures in brittle media and iterated conformal maps
Phys. Rev. E {\bf 65}, 045101 (2002).

\bibitem{bazant03} M. Z. Bazant, preprint, physics/0302086.

\bibitem{needham} T. Needham, {\it Visual Complex Analysis}
(Oxford, 1997).

%\bibitem{hughes} B. Hughes, {\it Random Walks and Random Environments}, Vol
%1 (Cambridge, 1995).

\bibitem{asympnote} D. Margetis and T. M. Squires (private communications).

%\bibitem{hinch} E. J. Hinch, {\it Perturbation Methods} (Cambridge, ???).


\end{thebibliography}
\end{document}